\documentclass[a4paper,12pt]{article}
\pdfoutput=1
\usepackage{amsfonts,amsthm,amsmath,amssymb,graphicx,hyperref,color,cite,tikz,tikz-3dplot}
\def\Tr{{\rm Tr\, }}

\newcommand{\be}{\begin{equation}}
\newcommand{\bea}{\begin{eqnarray}}

\newcommand{\ee}{\end{equation}}
\newcommand{\eea}{\end{eqnarray}}

\begin{document} 

{\LARGE{ \centerline{\bf Gravitational dynamics}}}  
{\LARGE{ \centerline{\bf from collective field theory}}}  

\vskip.5cm 

\thispagestyle{empty} 
\centerline{{\large\bf Robert de Mello Koch\footnote{{\tt robert@zjhu.edu.cn}} }}

\vspace{.8cm}
\centerline{{\it School of Science, Huzhou University, Huzhou 313000, China,}}

\vspace{.2cm}
\centerline{{\it School of Physics and Mandelstam Institute for Theoretical Physics,}}
\centerline{{\it University of the Witwatersrand, Wits, 2050, }}
\centerline{{\it South Africa }}

\vspace{1truecm}

%%%%%%%%%%%%%%%%%
\thispagestyle{empty}

\centerline{\bf ABSTRACT}

\vskip.2cm 
We consider the relevance of a collective field theory description for the AdS/CFT correspondence. Collective field theory performs a systematic reorganization of the degrees of freedom of a (non-gravitational) field theory, replacing the original loop expansion parameter $\hbar$ with $1/N$. Collective fields are over complete signalling a redundancy inherent in the theory. We propose that this over completeness is the mechanism by which one arrives at a holographic description, to be identified with the gravity dual. We find evidence for this by studying the redundancy of the collective field theory, showing that degrees of freedom in the bulk can be expressed as a linear combination of degrees of freedom contained in an arbitrarily small neighbourhood of the boundary. %In the dual quantum gravity this conclusion follows from the (constraints associated to the) diffeomorphism invariance of the theory.

\setcounter{page}{0}
\setcounter{tocdepth}{2}
\newpage
\tableofcontents
\setcounter{footnote}{0}
\linespread{1.1}
\parskip 4pt

{}~
{}~

\section{Introduction}

The discovery of holography \cite{tHooft:1993dmi,Susskind:1994vu,Maldacena:1997re} which states that gravitational dynamics in $d+1$ dimensions can be described by a $d$-dimensional non-gravitational theory, is an extremely important development in quantum gravity. To explore the mechanism behind holography, it is useful to ask how information localizes in a theory of quantum gravity \cite{Marolf:2008mf,Jacobson:2012ubm,Papadodimas:2012aq,Banerjee:2016mhh,Raju:2018zpn,Raju:2019qjq,Jacobson:2019gnm,Laddha:2020kvp,Chowdhury:2020hse,Raju:2020smc,Chowdhury:2021nxw,Raju:2021lwh}. The answer clarifies the physical origin of holography for gravitational theories as we now explain. 

In a canonical quantization of gravity the theory is described using wavefunctionals of the metric and matter fields on a given spatial slice. We restrict our discussion to spacetimes that are asymptotically AdS. In gravity, diffeomorphism invariance is a fundamental principle. It implies that not every wavefunctional is a physical state: a physical state must take the same value for configurations that can be related by a diffeomorphism that vanishes asymptotically. This requirement can be expressed in terms of a set of momentum and Hamiltonian constraints. It is clear that diffeomorphism invariance implies that the gravitational theory is highly redundant: many different wave functionals correspond to the same physical state. Perturbatively, the constraints expressing diffeomorphism invariance are enough to imply that two wave functionals that coincide at the boundary for an infinitesimal interval of time must coincide everywhere in the bulk \cite{Chowdhury:2021nxw}. In this way diffeomorphism invariance reproduces the hallmark feature of holography: the state in the bulk is completely determined by boundary data \cite{Chowdhury:2021nxw}. Said differently, once the constraints associated to diffeomorphism invariance are solved, the only degrees of freedom of the wavefunctional on a slice are its boundary values: once these boundary values are specified, the value of the wavefunctional in the bulk is determined uniquely by diffeomorphism invariance. To avoid any confusion we should remark that diffeomorphism invariance on its own is not enough to ensure that the theory is holographic and indeed, this result does not go through for the classical theory which is diffeomorphism invariant. As the arguments of \cite{Chowdhury:2021nxw} make clear, one also needs to use the fact that the energy is positive as well as the fact that energy eigenstates are not localized as a consequence of the uncertainty principle. Other discussions of the relationship between the bulk constraints arising from diffeomorphism invariance and holography, arrive at consonant conclusions\cite{Marolf:2008mf,Jacobson:2012ubm,Jacobson:2019gnm}.

A non-perturbative argument, reaching the same conclusion \cite{Laddha:2020kvp,Chowdhury:2020hse,Raju:2020smc,Raju:2021lwh} is also possible. In classical gravity, the Hamiltonian takes the form \cite{Regge:1974zd}
\bea
H&=&\int_{\Sigma}N_\mu C^\mu+H_\partial
\eea
where $N_\mu$ are the lapse and shift functions and $C^\mu$ are the first class constraints that generate diffeomorphisms. The term $H_\partial$ is a boundary term. When acting on a gauge invariant observable/state, the Hamiltonian is equal to a boundary term. Assume that the diffeomorphism invariance of classical general relativity extends to quantum gravity. Since the quantum wave function is invariant under bulk diffeomorphisms, it is reasonable to assume the Hamiltonian in the UV completed theory is still given by a boundary term. Boundary observables are gauge invariant and self-adjoint observables built from fields in the intersection of any neighbourhood of a bulk Cauchy slice with any neighbourhood of the boundary. The algebra generated by these operators is the boundary algebra associated with the given Cauchy slice. Since we are assuming that the Hamiltonian is given by a boundary term, the projector onto the vacuum state of the AdS gravity is an element of the boundary algebra and, with the assumption of the Reeh-Schleider property for the boundary algebra we learn that the boundary algebra is the full algebra of operators. This conclusion, that in a theory of quantum gravity, a copy of all the information available on a Cauchy slice is also available near the boundary of the Cauchy slice, is the principle of the holography of information \cite{Laddha:2020kvp,Chowdhury:2020hse,Raju:2020smc,Raju:2021lwh}. Diffeopmorphism invariance -- which is responsible for the enormous redundancy in the gravitational dynamics -- is a crucial input in reaching this conclusion. Again this is not a property of the classical theory since the Reeh-Schleider property was used and this relies on quantum entanglement. The idea just outlined is a concrete and explicit realization of {\it complementarity}: the idea that in quantum gravity degrees of freedom in one region can sometimes be equated to a combination of degrees of freedom in another region\cite{tHooft:1984kcu,Susskind:1993if,Papadodimas:2013jku,Papadodimas:2013wnh,Papadodimas:2015jra,Papadodimas:2015xma,Papadodimas:2012aq,Banerjee:2016mhh,Raju:2018zpn}. The holography of information tells us that every bulk degree of freedom is equal to some combination of boundary degrees of freedom.

The AdS/CFT correspondence \cite{Maldacena:1997re,Gubser:1998bc,Witten:1998qj}, which provides a concrete example of holography, claims that quantum gravity on AdS can be described using a conformal field theory (CFT) defined on the boundary of AdS. A systematic procedure which starts from the CFT and constructs the dual gravitational dynamics would provide a constructive understanding of the correspondence. Starting from a lower dimensional non-gravitational CFT we should construct a higher dimensional theory that is holographic. Concretely we should construct a higher dimensional theory whose degrees of freedom are highly redundant in such a way that all information available on a given Cauchy slice is also available near the boundary of the slice. Here we will argue that a systematic procedure to accomplish this already exists in the form of collective field theory developed in \cite{Jevicki:1979mb,Jevicki:1980zg} and we will suggest how the redundancy needed for gravity is built into the collective formalism.

The basic idea is simple to state. Collective field theory performs a reorganization of the degrees of freedom of the field theory so that the loop expansion parameter is $1/N$, as expected from AdS/CFT for the gravity dual. This reorganization of degrees of freedom is achieved in a very interesting way. The idea is to introduce a collective field, which is given by gauge invariants of the field variables. The collective field theory is then a systematic procedure to construct the dynamics of this collective field. An important feature of the collective field, central to this paper, is that it is over complete and therefore the collective field theory is a redundant description. The collective field is in general a multi-local field so it has a natural interpretation as a field in a higher dimensional spacetime. The collective formalism builds the theory as if there is an independent degree of freedom at each point in this higher dimensional spacetime, so that there is an enormous redundancy in the theory. This implies relations between degrees of freedom at different points in the higher dimensional space-time, very reminiscent of complementarity in gravity. We will give evidence that the redundancy of the collective field theory construction is exactly what is required to produce a holographic theory and so, which is required to produce a theory of gravity. Our evidence for this identification is obtained by using the operator product expansion of CFT to characterise the redundancy in the collective field. The collective field theory description of vector models is simple enough that we can carry this exercise out explicitly. In this example we argue that it is possible to reduce the theory to boundary degrees of freedom so that the collective field theory description of vector models is a holographic theory. Making the reasonable assumption that gravity is the only theory that is holographic\footnote{This is not proved but it is consistent with what we know about nature.} we conclude that collective field theory is a theory of gravity.  

In Section \ref{Collective} we review those aspects of collective field theory that are most relevant for our discussion. In particular, we emphasize that the collective field provides an over complete description and we argue that all degrees of freedom in this over complete set are being treated as independent. After this general discussion we turn to a detailed discussion of vector models in Section \ref{Bilocal}. In this case the collective fields are given by bilocal field variables and the dual gravity is given by Vassilliev's higher spin gravity \cite{Vasiliev:1990en,Vasiliev:2003ev}. There is a concrete holographic mapping between the CFT and the dual gravity, and using this mapping we are able to demonstrate our basic claim that the collective field theory is holographic. In Section \ref{matrixmodels} we briefly discuss non-Abelian gauge theories and give some evidence that the same mechanism for holography is relevant. We give some discussion of our results, conclusions and suggestions for future work in Section \ref{conclusions}.

\section{Collective Field Theory}\label{Collective}

In this section we review the collective field theory formalism which was developed in \cite{Jevicki:1979mb,Jevicki:1980zg}. Collective field theory performs a highly non-trivial reorganization of the degrees of freedom of a gauge theory, by changing from the original field variables to a new set of collective fields. In the collective description ${1\over N}$ is the loop expansion parameter. 

The collective fields are given by a set of over complete, commuting and gauge invariant variables. This over completeness of the collective fields, which is directly responsible for a redundancy in the collective dynamics, is a generic feature and it plays a central role in establishing the holographic nature of collective field theory. To make the discussion transparent, it is helpful to discuss concrete examples. Consider first the case of vector models. The field theory is defined using a field $\phi^a(t,\vec{x})$ with $a=1,2,\cdots,N$, which is an $O(N)$ vector. We consider a model defined in $d+1$-dimensional Minkowski spacetime. The Hamiltonian of the vector model is given by
\bea
H&=&\int \Bigl( -{1 \over 2}{\delta \over \delta \phi^a(t,\vec{x})}{\delta \over \delta \phi^a(t,\vec{x})} + {1 \over 2}\vec{\nabla}\phi^a(t,\vec{x})\cdot\vec{\nabla}\phi^a(t,\vec{x}) + v(\phi(t,\vec{x}) \cdot \phi(t,\vec{x})) \Bigr) d^d x\label{originalham}
\eea
where $v(\phi \cdot \phi)$ is any $O(N)$ invariant interaction. We quantize the field by imposing the usual equal time commutation relations
\bea
\left[ \pi^a(t,\vec{x}),\phi^b(t,\vec{y})\right]=-i\delta(\vec{x}-\vec{y})\delta^{ab}
\eea
This implies that the conjugate momentum can be written as
\bea
\pi^a (t,\vec{x})&=&{1\over i}{\delta\over\delta\phi^a(t,\vec{x})}
\eea
Declaring the $O(N)$ symmetry to be a gauge symmetry, the physical sector of the theory corresponds to the $O(N)$ singlet sector. In a Hamiltonian approach, the collective field theory provides a description of this sector by employing the equal time bilocal fields
\bea
\sigma (t,\vec{x}_1,\vec{x}_2)&=&\sum_{a=1}^N\phi^a(t,\vec{x}_1)\phi^a(t,\vec{x}_2)
\eea
The physical degrees of freedom of the original vector model are given by the singlet sector of $N$ scalar fields in $d+1$ dimensions. The above bilocal field is a single field defined in a $2d+1$ dimensional spacetime and it obeys a single constraint
\bea
\sigma (t,\vec{x}_1,\vec{x}_2)&=&\sigma (t,\vec{x}_2,\vec{x}_1)\label{SymCon}
\eea
Clearly we have many more degrees of freedom in $\sigma (t,\vec{x}_1,\vec{x}_2)$ than we have in $\phi^a (t,\vec{x})$ so we necessarily obtain a highly redundant description. Not all degrees of freedom in $\sigma (t,\vec{x}_1,\vec{x}_2)$ can possibly be independent. The collective field formalism \cite{Jevicki:1979mb,Jevicki:1980zg} ignores the over completeness of the collective fields, quantizing each degree of freedom independently. In particular, quantization is achieved by imposing the commutator\footnote{The two terms on the right hand side are needed to respect \eqref{SymCon}.}
\bea
\big[\Pi(t,\vec{x}_1,\vec{x}_2),\sigma(t,\vec{y}_1,\vec{y}_2)\big]=-i\delta(\vec{x}_1-\vec{y}_1)\delta(\vec{x}_2-\vec{y}_2)-i\delta(\vec{x}_1-\vec{y}_2)\delta(\vec{x}_2-\vec{y}_1)\label{cooquant}
\eea
to quantize the theory. Clearly we are quantizing many more degrees of freedom than are present in the original vector model. The commutator (\ref{cooquant}) implies the conjugate momentum
\bea
\Pi(t,\vec{x}_1,\vec{x}_2)&=&{1\over i}{\delta\over\delta \sigma(t,\vec{x}_1,\vec{x}_2)}
\eea
To derive the dynamics of the collective fields\footnote{We have not been precise about the detailed choice of gauge invariant variables that defines the collective field. A significant guiding principle is that this set must be chosen large enough that the application of the chain rule to the kinetic term in the Hamiltonian generates terms that again belong to the set.}, perform an operator change of variables \cite{Jevicki:1979mb} from $\phi^a(t,\vec{x})$ to the bilocal field $\sigma(t,\vec{x},\vec{y})$ using the chain rule
\bea
{\delta\over\delta\phi^a(t,\vec{x})}&=&\int d^d y \int d^d z\,{\delta\sigma(t,\vec{y},\vec{z})\over\delta\phi^a(t,\vec{x})}{\delta\over\delta\sigma(t,\vec{y},\vec{z})}
\eea
There is a non-trivial Jacobian associated with this change of variables which is determined by the requirement that the collective Hamiltonian is manifestly Hermittian. For more details the reader should consult \cite{Jevicki:1979mb}. Starting from the Hamiltonian (\ref{originalham}) the following equivalent representation in terms collective variables is obtained
\bea
H&=&2{\rm Tr}(\Pi \sigma \Pi)+{N^2 \over 8} \Tr \sigma^{-1} + \int d^d x\,\, v(\sigma(t,\vec{x},\vec{y}) \vert_{\vec{x}=\vec{y}}) \cr\cr
& &+{1 \over 2}\int d^d x\,\,\vec{\nabla}_{\vec{y}}\cdot\vec{\nabla}_{\vec{x}}\sigma(t,\vec{x},\vec{y})\Big|_{\vec{x}=\vec{y}}+\Delta V
\label{Hamcol}
\eea
$\Delta V$ summarizes ordering terms which are lower order in $1/N$ 
\bea
\Delta V&=&-{N \over 2}\Bigl( \int dx \delta(0) \Bigr){\rm Tr}\sigma^{-1}+{1 \over 2}\Bigl( \int dx \delta(0) \Bigr)^2 {\rm Tr} \sigma^{-1}\label{sublead}
\eea
In equations \eqref{Hamcol} and \eqref{sublead} the product of two bi-local fields is defined by
\bea
AB(t,\vec{x},\vec{z})&\equiv&\int d^d y\,\, A(t,\vec{x},\vec{y})\, B(t,\vec{y},\vec{z})
\eea
and the trace of a bi-local field is defined by
\bea
{\rm Tr}(A)=\int d^d x\,\,  A(t,\vec{x},\vec{x})
\eea
Given the over completeness of the description, one might question its validity. A proof that this description for vector models is correct, is the fact that it correctly generates the Schwinger-Dyson equations \cite{Jevicki:1979mb,Jevicki:1983hb} which determine the correlation functions of the invariant fields. Further, direct computation shows that the minimum of the collective potential
\bea
V_{\rm eff}&=&{N^2 \over 8} \Tr \sigma^{-1} + \int d^d x\,\, v(\sigma(t,\vec{x},\vec{y}) \vert_{\vec{x}=\vec{y}})
+{1 \over 2}\int d^d x\,\,\vec{\nabla}_{\vec{y}}\cdot\vec{\nabla}_{\vec{x}}\,\,\sigma(t,\vec{x},\vec{y})\Big|_{\vec{x}=\vec{y}}
\eea
generates the correct large $N$ gap equation, which is a non-perturbative (in $\hbar$ of the original field theory) result.

There is a second possible description of $O(N)$ vector models using collective field theory, which employs unequal time collective fields
\bea
\sigma (x_1^\mu,x_2^\mu)=\sum_{a=1}^N\phi^a(x_1^\mu)\phi^a(x_2^\mu)
\eea
which obey the single constraint
\bea
\sigma (x^\mu_1,x^\mu_2)&=&\sigma (x^\mu_2,x^\mu_1)
\eea
This is again manifestly an over complete description. The collective field theory in this case is determined \cite{Jevicki:1980zg} by performing a change of variables in the path integral description
\bea
\int D\phi^a(x^\mu)e^{iS}&=&\int D\sigma(x^\mu,y^\mu) \, J[\sigma]\,e^{iS[\sigma]}\,\,\equiv\,\,\int D\sigma(x^\mu,y^\mu) \,e^{iS_{\rm eff}}
\eea
The effective action is determined by the Jacobian of the change of variables. A straight forward way to determine this Jacobian \cite{Jevicki:1993rr} is by requiring that the Schwinger-Dyson equations derived in the original $\phi^a(x^\mu)$ variables agree with those derived using the $\sigma(x_1^\mu,x_2^\mu)$ variables\footnote{The set of gauge invariant variables that defines the collective field must be chosen large enough that the invariants appearing in the Schwinger-Dyson equations all belong to the set.}. The result is \cite{deMelloKoch:1996mj}
\bea
\ln J&=& (N-L^{d+1}\delta^{d+1}(0))\Tr\ln\sigma\qquad L^{d+1}=\int d^{d+1} x\qquad\delta^{d+1}(0)=\int {d^{d+1} p\over (2\pi)^{d+1}}
\eea
where the integral over $d^{d+1} p$ is an integral over momentum space. The equality of the resulting collective field theory dynamics with the original field theory dynamics can again be proved at the level of the Schwinger-Dyson equations. Explicit tests of the collective field theory include the demonstration that the saddle point of $S_{\rm eff}$ reproduces the usual large $N$ gap equation. Further, by working perturbatively it is easy to see that the collective field theory correctly reproduces the usual Feynman diagram expansion description of the  ${1\over N}$ expansion and subleading corrections in ${1\over N}$ come out correctly \cite{deMelloKoch:1996mj}.

We again want to stress the over completeness of this collective field description. One way to make sense of the path integral $\int D\phi^a(x^\mu)e^{iS}$ is by putting the theory on a lattice in which case the continuous coordinate $x^\mu$ is replaced by a discrete lattice label $\hat{i}$. The path integral becomes an integral over the value of the field at each lattice site $\phi(\hat{i})$. In this same approximation, the path integral over the collective field $\int D\sigma(x^\mu,y^\mu) \,e^{iS_{\rm eff}}$ would be an integral over the values of the collective field labelled by a pair of lattice sites $\sigma(\hat{i},\hat{j})$, with the constraint that $\sigma(\hat{i},\hat{j})=\sigma(\hat{j},\hat{i})$. The path integral involves many more integrals when expressed in terms of the collective field, than it did in the original variables. It is in this sense that collective field theory treats the degrees of freedom contained in the description using $\sigma(x_1^\mu,x_2^\mu)$ as independent. The over completeness of this description implies that the collective dynamics are necessarily redundant.

For matrix models it is more challenging to write down the relevant collective fields. This is a consequence of the fact that the set of gauge invariants in matrix models are much richer than the corresponding set for vector models. For matrix models, we can construct adjoint valued local operators at different spacetime locations and sew them together using open Wilson lines to produce a gauge invariant observable. Thus we will have multi-local collective fields, and not just bilocal collective fields. Even matrix quantum mechanics is challenging since for a multi-matrix model the set of gauge invariants is given by the trace of all words constructed using an alphabet with each different matrix giving a letter. A notable exception is the quantum mechanics of a single matrix $M_{ab}(t)$ for which the space of invariants is given by the trace of different powers of the matrix\footnote{In this case we can also use the eigenvalues to obtain a gauge invariant description.} $\Tr (M^n(t))$. In this example the collective field theory, using a Hamiltonian description, entails a change of variables from from the original matrix elements to the collective field given by
\bea
\phi (x,t)&=& \int_{-\infty}^{\infty} {dk\over 2\pi} e^{ikx}\Tr (e^{ikM})
\eea 
$\phi (x,t)$ is a scalar field in two dimensions, equal to the density of eigenvalues. The resulting collective field theory \cite{Das:1990kaa} gives an interacting theory with a cubic interaction. Detailed precision tests of this collective description show that it is the string field theory description of the $c=1$ string \cite{Demeterfi:1991tz,Demeterfi:1991nw,Jevicki:1991yi}. Notice that this description is again over complete: the single scalar field $\phi (x,t)$ has many more degrees of freedom than the original matrix $M_{ab}(t)$ has.

For the case of multi-matrix models it is possible to study the collective field theory numerically \cite{Jevicki:1982jj,Jevicki:1983wu,Koch:2021yeb,Mathaba:2023non}. These numerical results convincingly demonstrate that collective field theory gives a correct description of the dynamics of these models and that it has $1/N$ as loop expansion parameter. For this case too, the collective field theory description is over complete.

\section{Bilocal Holography for Vector Models}\label{Bilocal}

In the previous section we have explained that collective field theory gives a redundant description of the dynamics of a gauge theory, as a consequence of the fact that the collective field variables are over complete. This section aims to explore the details of this redundancy and for this purpose we return to the simple setting of vector models. We consider the free $O(N)$ vector model in 2+1 dimensions which is dual \cite{Klebanov:2002ja,Sezgin:2002rt} to higher spin gravity \cite{Vasiliev:1990en,Vasiliev:2003ev} in AdS$_4$. Collective field theory was first used to explore this duality in  \cite{Das:2003vw}. It was then further developed in a series of articles \cite{deMelloKoch:2010wdf,Jevicki:2011ss,Jevicki:2011aa,deMelloKoch:2012vc,deMelloKoch:2014mos,deMelloKoch:2014vnt,deMelloKoch:2018ivk,deMelloKoch:2021cni,deMelloKoch:2022sul,deMelloKoch:2023ngh} to achieve a detailed holographic mapping from the $2+1$ dimensional vector model CFT to the higher spin gravity in AdS$_4$. This mapping is constructed at large $N$ so that we obtain a free theory in the bulk of AdS$_4$. The key features of this mapping include
\begin{itemize}
\item[1.] A detailed mapping between the independent degrees of freedom of the CFT and the physical and independent degrees of freedom of the higher spin gravity. 
\item[2.] The bulk fields obey the correct equations of motion, with the correct boundary conditions so that we have a complete bulk reconstruction for all fields in the gravity theory.
\item[3.] The mapping gives a detailed account of what bulk fields can be reconstructed from a subregion of the CFT. This result reproduces expectations from entanglement wedge reconstruction.
\item[4.] Using this bilocal holography mapping, the statement of the holography of information for the gravitational degrees of freedom becomes the statement of the OPE for the CFT fields from which the bilocal is constructed.
\end{itemize}
The form of this dictionary is determined \cite{deMelloKoch:2010wdf} by the requirement that the reconstructed bulk fields obey the correct AdS$_4$ field transformations \cite{Metsaev:1999ui,Metsaev:2008fs}. Thus, the only input used to construct this mapping is conformal symmetry and the fact that it provides a demonstrably correct description of the dual gravity dynamics is significant. Using this map we can locate degrees of freedom in spacetime. This will be helpful when we explore the nature of the over completeness of the collective field and describe the redundancies in the collective field description. We will use capital letters to denote coordinates of the AdS$_4$ spacetime and little letters to denote the coordinates of the spacetime of the CFT$_3$. For a recent review of this material see \cite{deMelloKoch:2023ngh}.

For recent discussions of holography using the unequal time bilocal see \cite{Aharony:2020omh,Aharony:2021ovo,Aharony:2022feg} and for the holography of the IR fixed point see \cite{Mulokwe:2018czu,Johnson:2022cbe}.

\subsection{Lightfront Bilocal Holography Mapping}\label{light-frontmap}

The higher spin gravity has a single scalar field, as well as a single gauge field for every even integer spin $2s$. The gravity dual to the large $N$ limit of the CFT corresponds to free bulk fields. In this case we can use the Fronsdal description~\cite{Fronsdal:1978rb} rather than the full Vasiliev theory\cite{Vasiliev:1990en,Vasiliev:2003ev}. The spin-$2s$ Fronsdal field $A_{\mu_1 \mu_2 \cdots \mu_{2s}}$ is symmetric and obeys a double tracelessness condition $A{_\nu}{^\nu}{_\rho}{^\rho}{^{\mu_5\cdots \mu_s}}=0$. The AdS vierbein $e^A_\mu$ converts frame indices to spacetime indices. In the Poincar\'{e} patch of AdS we have
\bea
e^A_\mu = \frac{1}{z} \delta^A_\mu \, .
\eea
Denote Fronsdal fields with frame indices by $\Phi^{A_1 \cdots A_{2s}}$. Lightfront bilocal holography is obtained by completely gauge fixing and then reducing to independent field variables in gravity, and reducing to independent field variables in the conformal field theory. In the gravity theory we choose light cone gauge and solve the resulting constraints. In the end, only two components of the higher spin gauge field, at each spin, are physical and independent degrees of freedom. We choose these two components to be $\Phi^{XX\cdots XX}$ and $\Phi^{XX\cdots XZ}$ and collect the complete set of physical and independent fields into a single field, with the help of an additional variable $\theta$ as follows
\bea
\Phi(X^+,X^-,X,Z,\theta)=\sum_{s=0}^\infty \left(\cos (2s\theta) {\Phi^{XX\cdots XX}\over Z}+\sin (2s\theta){\Phi^{XX\cdots XZ}\over Z}\right)\label{thetaexpansion}
\eea
For what follows it is convenient to perform a Fourier transform to obtain
\bea
\Phi(X^+,P^+,X,Z,\theta)=\int \,dX^-\, e^{iP^+X^-} \Phi(X^+,X^-,X,\theta)
\eea
The conformal field theory dynamics is expressed as the collective field theory of an equal $x^+$ bilocal field
\bea
\sigma(x^+,x_1^-,x_1,x_2^-,x_2)=\sum_{a=1}^N\phi^a(x^+,x^-_1,x_1)\phi^a(x^+,x^-_2,x_2)
\eea
In formulating the map to the dual gravity theory, it is convenient to Fourier transform from $x^-$ to $p^+$ and work with the bilocal field
\bea
\sigma(x^+,p_1^+,x_1,p_2^+,x_2)&=&\int \, dx_1^-\,\int \,dx_2^-\,
e^{ip_1^+x_1^-+ip_2^+x_2^-}\sigma(x^+,x_1^-,x_1,x_2^-,x_2)\label{FTbi}
\eea
The holographic mapping between the CFT and higher spin gravity is now given by the following identification between the coordinates
\bea
x_1&=& X+Z \tan \left(\frac{\theta }{2}\right)\qquad
x_2\,\,=\,\, X-Z \cot \left(\frac{\theta }{2}\right)\qquad x^+=X^+\cr
p_1^+&=& P^+ \cos ^2\left(\frac{\theta }{2}\right)\qquad\quad
p_2^+\,\,=\,\, P^+ \sin ^2\left(\frac{\theta }{2}\right)\label{mapcft2grav}
\eea
and the fields
\bea
\Phi  &=& 2\pi P^+\sin\theta\,\, \eta\label{identifyfields}
\eea
The inverse of (\ref{mapcft2grav}) is
\bea
X&=& \frac{p_1^+ x_1+p_2^+ x_2}{p_1^++p_2^+}\qquad
Z\,\,=\,\,\frac{\sqrt{p_1^+ p_2^+} (x_1-x_2)}{p_1^++p_2^+}\cr
P^+&=& p_1^++p_2^+\qquad\qquad
\theta\,\,=\,\,2 \tan ^{-1}\left(\sqrt{\frac{p_2^+}{p_1^+}}\right)\label{mapgrav2cft}
\eea

\subsection{Covariant Bilocal Holography Mapping}\label{covariantmap}

The choice of lightcone gauge necessarily breaks the boundary Poincare invariance. By covariant bilocal holography we mean a mapping that preserves the boundary Poincare invariance. Covariant bilocal holography was developed in \cite{deMelloKoch:2023ngh}, relying heavily on results obtained in \cite{Metsaev:2008fs}. The higher spin gravity theory is formulated in the modified de Donder gauge \cite{Metsaev:2008ks} that leads to decoupled equations of motion. In this case we again collect the complete set of bulk higher spin gravity fields into a single field\footnote{We have suppressed indices on the $\phi_\pm^{(2s)}$ fields. These fields have $2s$ frame indices as in the discussion of the previous subsection.} that depends on an extra coordinate
\bea
\phi(x,Z,\varphi) &=&\sum_{s=0}^\infty \left({e^{i(2s-{1\over 2})\varphi}\over\sqrt{Z}}\phi_-^{(2s)}(x,Z)+{e^{-i(2s-{1\over 2})\varphi}\over\sqrt{Z}}\phi_+^{(2s)}(x,Z)\right)
\eea
Here $\phi_-^{(2s)}$ and $\phi_+^{(2s)}$ are the two physical polarizations of the higher spin gauge field of spin $2s$. The mapping between the coordinates is
\bea
X={p_1^0 x_1+p_2^0 x_2\over p_1^0+p_2^0} \qquad\qquad
Y={p_1^0 y_1+p_2^0 y_2\over p_1^0+p_2^0}\label{commap2grav1}
\eea
\bea
Z_1={\sqrt{p_1^0p_2^0}\over p_1^0+p_2^0}(x_1-x_2)\qquad\qquad
Z_2={\sqrt{p_1^0p_2^0}\over p_1^0+p_2^0}(y_1-y_2)\label{commap2grav2}
\eea
where $Z_1$ and $Z_2$ are related to $Z$ and $\varphi$ as follows
\bea
Z_1=Z\cos\varphi\qquad Z_2=Z\sin\varphi
\eea
When writing these formulas we have in mind bilocal fields composed of excitations that are described by wave packets tightly peaked about the energies $p_1^0$ and $p_2^0$. The above mapping is easily inverted
\bea
x_1=X+\sqrt{p_2^0\over p_1^0}Z_1\qquad x_2=X-\sqrt{p_1^0\over p_2^0}Z_1\label{commap2cft1}
\eea
\bea
y_1=Y+\sqrt{p_2^0\over p_1^0}Z_2\qquad y_2=Y-\sqrt{p_1^0\over p_2^0}Z_2
\label{commap2cft2}
\eea
The map between the fields is
\bea
\eta(t,X+\sqrt{p_2^0\over p_1^0}Z_1,Y+\sqrt{p_2^0\over p_1^0}Z_2,X-\sqrt{p_1^0\over p_2^0}Z_1,Y-\sqrt{p_1^0\over p_2^0}Z_2) =\phi(t,X,Y,Z_1,Z_2) \label{fieldident}
\eea

\subsection{Collective Field Theory is Holographic}

The collective field is over complete. The collective field theory formalism quantizes treating all degrees of freedom in the collective field as independent. Consequently collective field theory necessarily provides a redundant description. A natural consequence of this over completeness is that degrees of freedom in the collective field in one region can sometimes be equated to a combination of degrees of freedom of the collective field in another region. This is highly reminscent of the phenomenon of complementarity which arises in quantum gravity \cite{tHooft:1984kcu,Susskind:1993if,Papadodimas:2013jku,Papadodimas:2013wnh,Papadodimas:2015jra,Papadodimas:2015xma}. Making use of the mappings reviewed in Sections \ref{light-frontmap} and \ref{covariantmap} in this Section we will characterize the redundancy further, arguing that collective field theory is holographic. 

{\vskip 0.75cm}

\begin{figure}[h]%
\begin{center}
\begin{tikzpicture}
\draw[ultra thick] (0,0.75) -- (14,0.75);
%
%\draw[ultra thick] (2,0.75) arc (180:0:4 and 4);
\filldraw[red] (1,0.75) circle (0.15);
\filldraw[red] (2,0.75) circle (0.15);
\draw node at (1,0) {$\vec{x}_1$};
\draw node at (2,0) {$\vec{x}_2$};
\filldraw[blue] (9,0.75) circle (0.15);
\filldraw[blue] (10,0.75) circle (0.15);
\filldraw[blue] (11,0.75) circle (0.15);
\filldraw[blue] (12,0.75) circle (0.15);
\draw node at (9,0) {$\vec{x}_4$};
\draw node at (10,0) {$\vec{x}_5$};
\draw node at (11,0) {$\cdots$};
\draw node at (12,0) {$\vec{x}_n$};
\draw[->,thick] (2.1,3) -- (3.5,3);
\draw node at (6.5,3) {$X$ (parallel to the boundary)};
\draw[->,thick] (0,1.1) -- (0,3.5);
\draw node at (0.4,2.5) {$Z$};
\end{tikzpicture}
\caption{The horizontal direction, parametrized by $X$, is parallel to the boundary. The vertical direction, perpendicular to the boundary is parametrized by the emergent holographic coordinate $Z$. The fields within the bilocal $\eta(t,\vec{x}_1,\vec{x}_2)$ are at $\vec{x}_1$ and $\vec{x}_2$. The distance between these fields is much smaller than the distance to any other fields appearing in the correlator so that we can safely apply the OPE to $\eta(t,\vec{x}_1,\vec{x}_2)$ without worrying about convergence.}%
\label{justeta}%
\end{center}
\end{figure}
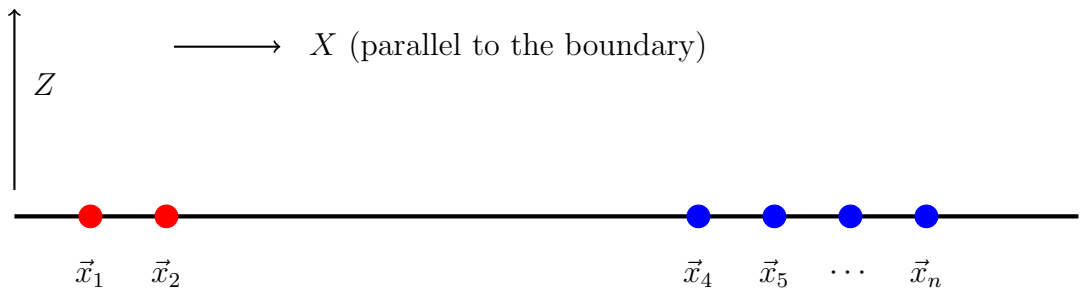

To start we will sketch our strategy in a simple example. Our basic tool to explore this redundancy is the operator product expansion (OPE). In conformal field theory the OPE expresses the product of two fields at different spacetime events as a sum of a (usually infinite) number of fields at a single event. The OPE is typically used within a correlator. The radius of convergence of the OPE inside a correlator is not predetermined but depends on the next-closest operator insertion. Consider a correlator of the form
\bea
\langle 0|\cdots\,\,\eta(t,\vec{x}_1,\vec{x}_2)|0\rangle\label{tosimplify}
\eea
where $\cdots$ stands for the remaining operators in the correlator and the separation between $|\vec{x}_1-\vec{x}_2|$ is small compared to the distance to any other operator. See Fig \ref{justeta} for an illustration. In this case the OPE can be applied to the bilocal field which itself is given by a product of two scalar fields. This implies a relation between a bilocal with separated fields and a bilocal with coincident fields. Since the OPE replaces the bilocal collective field with a sum of single trace primaries of the CFT, knowing where in the bulk the original bilocal maps to as well as where in the bulk the single trace primaries map to, we find an equality between fields that are defined at different locations. With the basic idea stated, we will now fill in the details.

We need the OPE of two free scalar fields which is given by (see for example \cite{deMelloKoch:2022sul})
\bea
\sum_{a=1}^N:\phi^a(x^\mu+y^\mu)\phi^a(x^\mu-y^\mu):&=&
\sum_{s=0}^\infty\sum_{d=0}^\infty c_{sd} 
\left(y^\mu {\partial\over\partial x^\mu}\right)^{2d} J_{2s}(y,x)
\label{explicitOPE}
\eea 
The sum on the right hand side goes over the complete set of single trace primary operators, which includes a scalar $J_0$ and a conserved current of every even integer spin $J_{2s}(y,x)$. The number $c_{sd}$ tells us about the contribution of the level $2d$ descendant of the primary current with spin $2s$. The explicit expressions for these coefficients are
\bea
c_{0d}={1\over 2^{2d} (d!)^2} \qquad {\rm and}\qquad
c_{sd}={(2 s)! (4 s-1)!!\over d! 2^{2 d+4 s-1} (d+2 s)!}\qquad\qquad s>0
\eea
The explicit formula for the conserved currents is
\begin{eqnarray}
J_s(y,x)
&=&J_{\mu_1\mu_2\cdots\mu_s}(x)y^{\mu_1}y^{\mu_2}\cdots y^{\mu_s}\cr\cr
&=&\sum_{k=0}^{s}
\frac{(-1)^k\, :(y\cdot\partial)^{s-k}\phi^a(x)\;(y\cdot\partial)^k \phi^a(x) :}
{k!(s-k)!\Gamma(k+{1\over 2})\Gamma(s-k+{1\over 2})}\label{scurrent}
\end{eqnarray}
Since the result of evaluating the OPE is expressed in terms of a sum over single trace primaries, it is interesting to ask where these single trace primaries map to in the AdS$_4$ bulk. Using the covariant description, the scalar primary is given by
\bea
\phi^a(t,\vec{x})\phi^a(t,\vec{x})=\eta(t,\vec{x},\vec{x})
\eea
Although this field contains many different $p^0$ contributions, by \eqref{commap2grav2} it is clear that these all map to $Z=0$. Exactly the same conclusion follows if we use the lightfront bilocal holography map. To express the conserved currents \eqref{scurrent} in terms of the bilocal field, we need to separate the locations of the fields slightly so that we can act with derivatives on either field separately. For example, in the covariant description we have
\begin{eqnarray}
J_s(x^\mu_1,y)
&=&\sum_{k=0}^{s}
\frac{(-1)^k\, :(y\cdot\partial_1)^{s-k}\;(y\cdot\partial_2)^{k} :}
{k!(s-k)!\Gamma(k+{1\over 2})\Gamma(s-k+{1\over 2})}
\eta(t,\vec{x}_1,\vec{x}_2)\Big|_{\vec{x}_2=\vec{x}_1}
\cr\cr
&\equiv& D^{(s)}\eta (t,\vec{x}_1,\vec{x}_1)
\end{eqnarray}
It is enough to separate $\vec{x}_1$ and $\vec{x}_2$ by an arbitrarily small amount $\epsilon=|\vec{x}_1-\vec{x}_2|$, evaluate the relevant derivatives and then set $\vec{x}_2=\vec{x}_1=\vec{x}$. Thus, we can construct the current at $\vec{x}$ from the bilocal field $\eta(t,\vec{x}_1,\vec{x}_2)$ with $|\vec{x}_1-\vec{x}_2|<\epsilon$ where $\epsilon$ is arbitrarily small. From \eqref{scurrent} we have
\bea
Z={\sqrt{p_1^0 p_2^0}\over p_1^0+p_2^0}|\vec{x}_1-\vec{x}_2|
\eea
Since the energy $p^0>0$, we know that $0<{\sqrt{p_1^0p_2^0}\over p_1^0+p_2^0}<1$. Consequently, the conserved currents all map to an arbitrarily small neighbourhood of the boundary $Z<\epsilon$. Precisely the same conclusion is easily demonstrated using the lightcone bilocal map. We conclude that the complete set of single trace primary operators, after mapping to the dual gravity, are supported in an arbitrarily small neighbourhood of the boundary. Returning to \eqref{tosimplify} we can write
\bea
\langle 0|\cdots\,\,\eta(t,\vec{x}_1,\vec{x}_2)|0\rangle&=&
\sum_{s=0}^\infty\sum_{d=0}^\infty \, c_{sd}\,\left((x_1-x_2)^\mu {\partial\over\partial x_1^\mu}\right)^{2d}\,\langle 0|\cdots\,\,D^{(2s)}\eta (t,\vec{x}_1,\vec{x}_1)|0\rangle
\eea
By taking $|\vec{x}_1-\vec{x}_2|$ to be arbitrarily well separated $\eta(t,\vec{x}_1,\vec{x}_2)$ on the LHS maps to a higher spin gravity field located arbitrarily deep in the bulk while $D^{(2s)}\eta (t,\vec{x}_1,\vec{x}_1)$ is located in an arbitrarily small neighbourhood of the boundary. Thus, under the conditions stated above, the redundancy in the collective field description implies that any bulk field can be written as a linear combination of fields located at the boundary.

{\vskip 0.75cm}

\begin{figure}[h]%
\begin{center}
\begin{tikzpicture}
\draw[ultra thick] (0,0.75) -- (15,0.75);
%
%\draw[ultra thick] (2,0.75) arc (180:0:4 and 4);
\filldraw[red] (1,0.75) circle (0.15);
\filldraw[red] (1.75,0.75) circle (0.15);
\filldraw[red] (3.25,0.75) circle (0.15);
\filldraw[red] (4,0.75) circle (0.15);
\draw node at (1,0) {$\vec{x}_1$};
\draw node at (1.75,0) {$\vec{x}_3$};
\draw node at (3.25,0) {$\vec{x}_2$};
\draw node at (4,0) {$\vec{x}_4$};
\filldraw[blue] (11,0.75) circle (0.15);
\filldraw[blue] (12,0.75) circle (0.15);
\filldraw[blue] (13,0.75) circle (0.15);
\filldraw[blue] (14,0.75) circle (0.15);
\draw node at (11,0) {$\vec{x}_4$};
\draw node at (12,0) {$\vec{x}_5$};
\draw node at (13,0) {$\cdots$};
\draw node at (14,0) {$\vec{x}_n$};
\draw[->,thick] (2.1,3) -- (3.5,3);
\draw node at (6.5,3) {$X$ (parallel to the boundary)};
\draw[->,thick] (0,1.1) -- (0,3.5);
\draw node at (0.4,2.5) {$Z$};
\end{tikzpicture}
\caption{The fields within the bilocal $\eta(t,\vec{x}_1,\vec{x}_2)$ are at $\vec{x}_1$ and $\vec{x}_2$ and the fields within the bilocal $\eta(t,\vec{x}_3,\vec{x}_4)$ are at $\vec{x}_3$ and $\vec{x}_4$. The convergent OPE channel is when we first collapse the fields at $\vec{x}_1$ and $\vec{x}_3$ to obtain a sum over operators at ${\vec{x}_1+\vec{x}_3\over 2}$ and we collapse the fields at $\vec{x}_2$ and $\vec{x}_4$ to obtain a sum over operators at ${\vec{x}_2+\vec{x}_4\over 2}$. Finally we use the OPE to collapse operators at ${\vec{x}_1+\vec{x}_3\over 2}$ and operators at ${\vec{x}_2+\vec{x}_4\over 2}$}%
\label{etaeta}%
\end{center}
\end{figure}
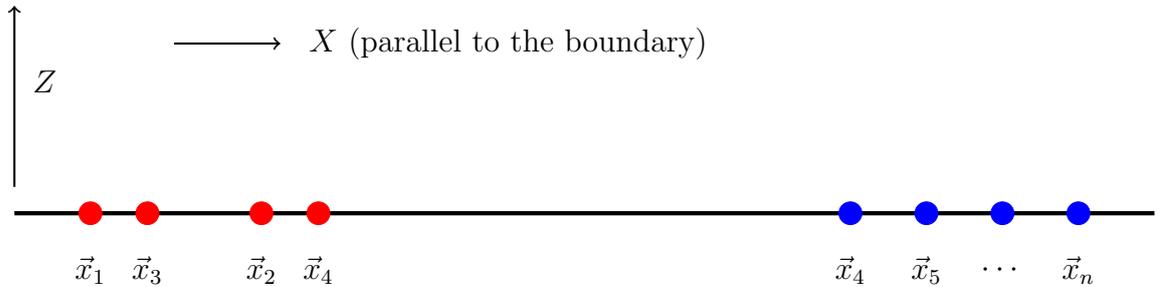

Although this is highly suggestive, it is not yet enough to conclude that collective field theory is holographic. The point is that there are many different channels for the OPE i.e. many different choices about the order in which pairs of fields are collapsed to a local field using the OPE. This is not determined by the way color indices $a$ are contracted, but rather it is determined by the location of the fields in spacetime. As an example, we might have the correlator
\bea
\langle 0|\cdots\,\,\eta(t,\vec{x}_1,\vec{x}_2)\eta(t,\vec{x}_3,\vec{x}_4)|0\rangle\label{tosimplify2}
\eea
Here $\cdots$ again stands for additional operators in the correlators and the distances $l_{ij}=|\vec{x}_i-\vec{x}_j|$ for $i,j\in\{1,2,3,4\}$ are small compared to the distance to any operator in $\cdots$. In the case that $l_{13}$ and $l_{24}$ are small compared to the other distances $l_{ij}$, the convergent OPE channel is obtained by collapsing the pair of scalar fields at $\vec{x}_1$ and $\vec{x}_3$ producing a sum over operators at $\vec{x}_1$, and the pair at $\vec{x}_2$ and $\vec{x}_4$ producing a sum over operators at $\vec{x}_2$. Since the pairs of fields to which the OPE was applied do not correspond to a single bilocal, the result of these two applications of the OPE does not produce an $O(N)$ singlet. An $O(N)$ singlet is however obtained if we now use the OPE again to collapse the fields at $\vec{x}_1$ with those at $\vec{x}_2$. We will now make the plausible assumption that the algebra of the single trace primaries generates the complete set of gauge invariant operators. Under this assumption, since the local single trace primaries map to operators defined in an arbitrarily small neighbourhood of the boundary, the complete set of local gauge invariant operators will again map to operators that belong to the same neighbourhood. Consequently, in the above example, the product of collective fields $\eta(t,\vec{x}_1,\vec{x}_2)\eta(t,\vec{x}_3,\vec{x}_4)$ which corresponds to a product of two fields in the bulk, can again be replaced by a linear combination of fields located at the boundary.

It is clear how this argument generalizes to the product of an arbitrary number of $\eta$ fields, so that we can conclude that the redundancy present in the collective field theory is such that an arbitrary product of bulk operators can be replaced by a linear combination of fields located at the boundary. We therefore conclude that collective field theory is holographic and so it is a theory of gravity.

\subsection{Collective Redundancy and Diffeomorphism Invariance}

Diffeomorphism invariance is a fundamental principle of gravity. It ensures that the theory doesn't depend on the specific coordinates used to describe spacetime and consequently that it is a geometric theory. Diffeomorphism invariance is realized in the theory as a gauge symmetry, so that it signals a huge redundancy in the description of gravity: two states related by a diffeomorphism correspond to the same physical state. Reducing to the non-redundant degrees of freedom we find that the theory is holographic \cite{Chowdhury:2021nxw}. This is a rather beautiful conclusion: endowing a theory with diffeomorphism invariance is a way in which we can ensure that it is holographic.

A central feature of collective field theory is that it uses an over complete set of collective fields to describe the CFT. Each degree of freedom in the collective field is treated as independent so that the collective field theory has more degrees of freedom that the original CFT had. Collective field theory is therefore necessarily a redundant description. In the previous section we have studied this redundancy, arguing that collective field theory is holographic.

These results suggest that perhaps the redundancy present in collective field theory is the origin of diffeomorphism invariance of the gravitational theory that is the AdS/CFT dual of the CFT. Indeed, if this is not the case it means that collective field theory is a completely different way in which holographic theories can be constructed, which seems far less plausible. Recall that the constraints of diffeomorphism invariance are summarized in the Hamiltonian and momentum constraints. The integrated Hamiltonian constraint, which played a central role in the analysis of \cite{Chowdhury:2021nxw}, leads to the gravitational Gauss law. It basically states an equality of a bulk energy and the boundary ADM Hamiltonian. A very similar statement can be derived in collective field theory by the following logic: consider a collective field which maps to a gravity field in the bulk of AdS. We can evolve it through an infinitesimal time interval by computing it's commutator with the Hamiltonian, and then use the OPE to map the result to the boundary. Alternatively, we could first use the OPE to map this to a combination of boundary degrees of freedom and then evolve it through an infinitesimal time interval using the Hamiltonian. The equality of these two procedures is the statement that the dynamics respects the redundancy of the collective field and it is assured by the consistency of collective field theory. In this way we relate the action of the Hamiltonian on bulk degrees of freedom to the action of the Hamiltonian on boundary degrees of freedom. It would be interesting to see if the gravitational Gauss law can be reproduced in this way.

The gravitational Gauss law is obtained by integrating the Hamiltonian constraint. The Hamiltonian and momentum constraints are an infinite number of constraints, holding at each point of the Cauchy slice. The solution to the integrated Hamiltonian constraint does not itself give a solution to these constraints at each point. However, \cite{Chowdhury:2021nxw} have argued that each solution of the gravitational Gauss law can be uniquely uplifted to a solution of the pointwise constraints. To be more explicit, the fluctuation of the metric about the AdS background can be decomposed into a transverse-traceless (TT), a longitudinal (L) and a T component (see \cite{Chowdhury:2021nxw} for details)
\bea
h_{ij}&=&h_{ij}^{TT}+h_{ij}^L+h_{ij}^T
\eea
The solution to the gravitational Gauss law fixes the dependence of the wavefunctional on $h_{ij}^{TT}$ and the matter fields. Uplifting the solution determines the dependence of the wavefunctional on $h_{ij}^L$ and $h_{ij}^T$. This analysis suggests that there is a consistent truncation of the dynamics to $h_{ij}^{TT}$ and the matter fields. It is natural to expect that this truncation of the gravity corresponds to the collective field theory description obtained using the equal time bilocals\footnote{The wavefunctionals of $h^{TT}$ and the matter obtained in \cite{Chowdhury:2021nxw} employ an ordinary free-field Fock space. The large $N$ equal time bilocals are also described using an ordinary free-field Fock space.}, while the description before truncation corresponds to the description obtained using unequal time bilocals. Of course, the gravity dual to the bilocals has to be supplemented with higher spin gauge fields, representing a non-trivial extension of the analysis of \cite{Chowdhury:2021nxw}. 

\subsection{Additional redundancies at finite $N$}

The redundancies we have exhibited above arise because the collective field is a composite of fundamental fields. Since the fundamental fields themselves obey a non-trivial algebra, there are non-trivial relations between the bilocal field. We have explored these redundancies using the OPE. The structure and properties of the OPE are largely determined by conformal symmetry and are not dependent on the value of $N$. In particular, these redundancies are present in the large $N$ limit and we expect them to be present in perturbative gravity. It is then possible that these match the redundancies that are exhibited in a perturbative analysis of the Wheeler-De Witt equation.

There are additional redundancies that appear for finite $N$. Since $1/N$ sets the strength of the gravitational interaction, these effects correspond to the physics of strong gravity fields. The simplest way to exhibit these redundancies is at $N=1$ where the bilocal is just the product of two fields
\bea
\sigma(t,\vec{x},\vec{y})=\phi(t,\vec{x})\phi(t,\vec{y})
\eea
We obviously have the identities
\bea
\sigma(t,\vec{x}_1,\vec{x}_2)\sigma(t,\vec{x}_3,\vec{x}_4)=\sigma(t,\vec{x}_1,\vec{x}_3)\sigma(t,\vec{x}_2,\vec{x}_4)=\sigma(t,\vec{x}_1,\vec{x}_4)\sigma(t,\vec{x}_2,\vec{x}_3)
\eea
These are again because the field is a composite, but they are of course, not related to conformal symmetry. For the OPE we had to carefully consider the conditions under which the OPE converges. That is not the case for the identities above. To see the redundancies present for a general but finite value of $N$, choose a point $\vec{x}$ and a collection of $N+1$ points $\vec{y}_i$, $i=1,2,\cdots N+1$. It is then straightforward to verify that
\bea
\epsilon^{i_1 i_2\cdots i_{N+1}}\sigma(t,\vec{y},\vec{x}_{i_1})\sigma(t,\vec{y},\vec{x}_{i_2})\cdots \sigma(t,\vec{y},\vec{x}_{i_{N+1}})=0
\eea
The product of operators given above creates a state of dimension $N+1$ and hence this redundancy is a new redundancy that is present between heavy operators in the dual gravity theory. These redundancies were studied in vector models at finite temperature in \cite{Shenker:2011zf}. There is a large $N$ transition at a very high temperature of order $\sqrt{N}$ driven by the decrease in the number of degrees of freedom from that of the simple higher spin gas, due to these relations.

\section{Non-Abelian Gauge Theories}\label{matrixmodels}

Our discussion has focused on vector models. Vector models are simple enough that we can explicitly construct the holographic mapping between the CFT and the dual AdS gravity. This simplicity follows because the complete set of single trace gauge invariant operators can conveniently be packaged in a single bilocal field, making the bilocal a natural candidate for the collective field. For matrix models involving more than one matrix, the space of gauge invariant observables is much richer. Although one can write down candidate collective fields, working with these fields is considerably more difficult and constructing the holographic mapping in this case remains an important outstanding problem. However, as we discuss below, even without the detailed holographic map there is evidence suggesting that the argument given above for vector models will generalize to matrix models.

The collective field is constructed, as usual, from the set of gauge invariant fields. To construct gauge invariant operators we can use Wilson lines to sew together local adjoint valued local operators that are located at different spacetime locations. Our arguments in Section \ref{Bilocal} above used the detailed mapping between gauge invariant operators in the CFT and operators in the bulk of the dual gravity. Although we can't be precise about these details, we do have some general expectations based on our experience with AdS/CFT. The basic observation we use is the scale radius duality \cite{Susskind:1998dq}, which tells us that point like operators are located at the boundary of AdS, whereas extended operators are located deep in the bulk. Thus, scale/radius duality suggests that the holographic mapping will place local operators at the boundary of AdS and extended operators deeper in the bulk. The larger the ``size'' of the operator, the deeper it sits in the bulk. This is very obviously a property of the vector model mapping given in Sections \ref{light-frontmap} and \ref{covariantmap}, where the distance between the two fields in the bilocal does indeed determine the value of the $Z$ coordinate. Appealing to scale/radius duality, we will assume that this will also be a property of the holographic mapping for multi-matrix CFTs like ${\cal N}=4$ super Yang-Mills theory.

In this case we can again use the OPE to collapse a product of operators at different locations into a local operator. With our assumption about the holographic mapping for multi-matrix CFTs, the OPE again relates bulk degrees of freedom to a linear combination of boundary degrees of freedom. This shows that the collective field theory description of multi-matrix CFTs is a redundant description and that the theory is again holographic i.e. that the collective field theory of multi-matrix CFTs is a theory of gravity.

Just as is the case for vector models, in matrix models we also expect additional redundancies to arise at finite $N$. These relations are between composite operators and they first appear when the composite operators contain more than $N$ fields. Thus this redundancy is again a statement about the equality of heavy operators in the dual gravity theory. Bases for these operators can be given by operators labelled by Young diagrams \cite{Corley:2001zk,Brown:2007xh,Bhattacharyya:2008rb}, in which case these finite $N$ relations are solved by restricting to operators labelled by Young diagrams with no more than $N$ rows. These finite $N$ redundancies are again very different to redundancies that are exhibited by using the OPE.

\section{Conclusions}\label{conclusions}

This article has discussed the collective field theory description of vector models and multi-matrix models, which are both ordinary (non-gravitational) field theories. For the case of a single matrix, it's well known that the collective field theory of a single Hermittian matrix leads to the Das-Jevicki Hamiltonian \cite{Das:1990kaa} which reproduces the string field theory of the $c=1$ string. For the vector model the collective description leads to bilocal holography, which gives an explicit and detailed map between the CFT and the dual higher spin gravity. In both of these cases, collective field theory leads to a description in terms of gravitational dynamics. In this article we have put forward an explanation of the origin of that gravitational dynamics. The hallmark of gravity is diffeomorphism invariance, which signals an enormous redundancy in the dynamics of the system. Solving the constraints implied by diffeomorphism invariance and thereby reducing to the non-redundant degrees of freedom shows that these theories are holographic: the number of physical degrees of freedom grows like the area of the boundary of the system. The collective field theory also has an enormous redundancy in its dynamics. This redundancy appears because the collective field is in fact an over complete description, and the collective field formalism treats each degree of freedom in the collective field as independent. We have argued that the redundancy introduced by the over complete collective field is precisely what is needed to produce a holographic theory i.e. a theory of gravity. Our evidence for this was arrived at by carefully studying this redundancy in the case of bilocal holography and concluding that any bulk field can be expressed as a combination of boundary degrees of freedom. For the case of the collective field description of multi-matrix models CFTs, we have given some plausibility arguments that the same conclusion holds.

We have speculated that there may be a direct relation between the redundancy in the collective field theory description and diffeomorphism invariance in the dual gravity theory. As we have argued above, applying the OPE to the collective field exhibits an equality between naively independent degree of freedom. The dynamics of collective field theory, which respects conformal invariance, will respect identifications deduced using the OPE. We can rewrite a field using the OPE and then evolve it dynamically, or we could evolve it dynamically and then rewrite it using the OPE. These two procedures must give the same result. This suggests that one way in which the dynamical content of the collective redundancy could be approached is to ask how the dynamics is constrained by the requirement that it commutes with the OPE, as just described.

It is worth noting that the collective field theory formalism is equally applicable to ordinary quantum field theories as well as the more special case of conformal field theories. For the collective field theory of the Hermitian matrix model \cite{Das:1990kaa} the theory is not conformally invariant. In this case the collective field theory reproduces the $c=1$ string so it represents an example of gauge theory/gravity duality which is not obviously implied by the AdS/CFT  correspondence. For vector models, the redundancy is present regardless of whether the theory is conformal. The conformal symmetry does however make it simple (through the use of the OPE) to analyse the redundancies present in the collective field. It would be interesting to further explore the gauge/gravity duality for these non-conformal quantum field theories, as well as for multi-matrix theories. For the case of multi-matrix theories, deriving the holographic mapping for the simpler case of a conformal field theory would be significant progress.

As a final remark, our results point out an interesting interplay between the gauge symmetry of the original CFT and gravity dynamics: the original gauge symmetry determines the form of the gauge invariant fields and hence of the collective field. The collective field is redundant and this redundancy is highly dependent on details of this gauge symmetry. The holographic nature of the theory then arises from the redundancy in this description.

\begin{center} 
{\bf Acknowledgements}
\end{center} 
This research is supported by a start up research fund of Huzhou University, a Zhejiang Province talent award and by a Changjiang Scholar award. The author would like to thank the Isaac Newton Institute for Mathematical Schiences for support and hospitality during the programme ``Black holes: bridges between number theory and holographic quantum information'' when work on this paper was completed. This work was supported by EPSRC Grant Number EP/R014604/1. We thank Suvrat Raju for helpful correspondence on \cite{Chowdhury:2021nxw}. We thank Cameron Beetar, Garry Kemp, Jaco Van Zyl and especially Antal Jevicki for very useful discussions on the subject of this paper.


\begin{thebibliography}{}

\bibitem{tHooft:1993dmi}
G.~'t Hooft, ``Dimensional reduction in quantum gravity,'' Conf. Proc. C \textbf{930308} (1993), 284-296 [arXiv:gr-qc/9310026 [gr-qc]].

\bibitem{Susskind:1994vu}
L.~Susskind, ``The World as a hologram,'' J. Math. Phys. \textbf{36} (1995), 6377-6396
doi:10.1063/1.531249 [arXiv:hep-th/9409089 [hep-th]].

\bibitem{Maldacena:1997re}
J.~M.~Maldacena, ``The Large N limit of superconformal field theories and supergravity,''
Adv. Theor. Math. Phys. \textbf{2} (1998), 231-252 doi:10.4310/ATMP.1998.v2.n2.a1
[arXiv:hep-th/9711200 [hep-th]].

\bibitem{Marolf:2008mf}
D.~Marolf, ``Unitarity and Holography in Gravitational Physics,'' Phys. Rev. D \textbf{79} (2009), 044010 doi:10.1103/PhysRevD.79.044010 [arXiv:0808.2842 [gr-qc]].

\bibitem{Jacobson:2012ubm}
T.~Jacobson, ``Boundary unitarity and the black hole information paradox,'' Int. J. Mod. Phys. D \textbf{22} (2013), 1342002 doi:10.1142/S0218271813420029 [arXiv:1212.6944 [hep-th]].

\bibitem{Papadodimas:2012aq}
K.~Papadodimas and S.~Raju, ``An Infalling Observer in AdS/CFT,'' JHEP \textbf{10} (2013), 212 doi:10.1007/JHEP10(2013)212 [arXiv:1211.6767 [hep-th]].

\bibitem{Banerjee:2016mhh}
S.~Banerjee, J.~W.~Bryan, K.~Papadodimas and S.~Raju, ``A toy model of black hole complementarity,'' JHEP \textbf{05} (2016), 004 doi:10.1007/JHEP05(2016)004 [arXiv:1603.02812 [hep-th]].

\bibitem{Raju:2018zpn}
S.~Raju, ``A Toy Model of the Information Paradox in Empty Space,'' SciPost Phys. \textbf{6} (2019) no.6, 073 doi:10.21468/SciPostPhys.6.6.073 [arXiv:1809.10154 [hep-th]].

\bibitem{Raju:2019qjq}
S.~Raju, ``Is Holography Implicit in Canonical Gravity?,'' Int. J. Mod. Phys. D \textbf{28} (2019) no.14, 1944011 doi:10.1142/S0218271819440115 [arXiv:1903.11073 [gr-qc]].

\bibitem{Jacobson:2019gnm}
T.~Jacobson and P.~Nguyen, ``Diffeomorphism invariance and the black hole information paradox,'' Phys. Rev. D \textbf{100} (2019) no.4, 046002 doi:10.1103/PhysRevD.100.046002
[arXiv:1904.04434 [gr-qc]].

\bibitem{Laddha:2020kvp}
A.~Laddha, S.~G.~Prabhu, S.~Raju and P.~Shrivastava, ``The Holographic Nature of Null Infinity,'' SciPost Phys. \textbf{10} (2021) no.2, 041 doi:10.21468/SciPostPhys.10.2.041
[arXiv:2002.02448 [hep-th]].

\bibitem{Chowdhury:2020hse}
C.~Chowdhury, O.~Papadoulaki and S.~Raju, ``A physical protocol for observers near the boundary to obtain bulk information in quantum gravity,'' SciPost Phys. \textbf{10} (2021) no.5, 106 doi:10.21468/SciPostPhys.10.5.106 [arXiv:2008.01740 [hep-th]].

\bibitem{Raju:2020smc}
S.~Raju, ``Lessons from the information paradox,'' Phys. Rept. \textbf{943} (2022), 1-80
doi:10.1016/j.physrep.2021.10.001 [arXiv:2012.05770 [hep-th]].

\bibitem{Chowdhury:2021nxw}
C.~Chowdhury, V.~Godet, O.~Papadoulaki and S.~Raju, ``Holography from the Wheeler-DeWitt equation,'' JHEP \textbf{03} (2022), 019 doi:10.1007/JHEP03(2022)019 [arXiv:2107.14802 [hep-th]].

\bibitem{Raju:2021lwh}
S.~Raju, ``Failure of the split property in gravity and the information paradox,'' Class. Quant. Grav. \textbf{39} (2022) no.6, 064002 doi:10.1088/1361-6382/ac482b [arXiv:2110.05470 [hep-th]].

\bibitem{Regge:1974zd}
T.~Regge and C.~Teitelboim, ``Role of Surface Integrals in the Hamiltonian Formulation of General Relativity,'' Annals Phys. {\bf 88} (1974), 286 doi:10.1016/0003-4916(74)90404-7.

\bibitem{tHooft:1984kcu}
G.~'t Hooft, ``On the Quantum Structure of a Black Hole,'' Nucl. Phys. B \textbf{256} (1985), 727-745 doi:10.1016/0550-3213(85)90418-3

\bibitem{Susskind:1993if}
L.~Susskind, L.~Thorlacius and J.~Uglum, ``The Stretched horizon and black hole complementarity,'' Phys. Rev. D \textbf{48} (1993), 3743-3761 doi:10.1103/PhysRevD.48.3743 [arXiv:hep-th/9306069 [hep-th]].

\bibitem{Papadodimas:2013jku}
K.~Papadodimas and S.~Raju, ``State-Dependent Bulk-Boundary Maps and Black Hole Complementarity,'' Phys. Rev. D \textbf{89} (2014) no.8, 086010 doi:10.1103/PhysRevD.89.086010 [arXiv:1310.6335 [hep-th]].

\bibitem{Papadodimas:2013wnh}
K.~Papadodimas and S.~Raju, ``Black Hole Interior in the Holographic Correspondence and the Information Paradox,'' Phys. Rev. Lett. \textbf{112} (2014) no.5, 051301 doi:10.1103/PhysRevLett.112.051301 [arXiv:1310.6334 [hep-th]].

\bibitem{Papadodimas:2015jra}
K.~Papadodimas and S.~Raju, ``Remarks on the necessity and implications of state-dependence in the black hole interior,'' Phys. Rev. D \textbf{93} (2016) no.8, 084049
doi:10.1103/PhysRevD.93.084049 [arXiv:1503.08825 [hep-th]].

\bibitem{Papadodimas:2015xma}
K.~Papadodimas and S.~Raju, ``Local Operators in the Eternal Black Hole,'' Phys. Rev. Lett. \textbf{115} (2015) no.21, 211601 doi:10.1103/PhysRevLett.115.211601 [arXiv:1502.06692 [hep-th]].

\bibitem{Gubser:1998bc}
S.~S.~Gubser, I.~R.~Klebanov and A.~M.~Polyakov, ``Gauge theory correlators from noncritical string theory,'' Phys. Lett. B \textbf{428} (1998), 105-114 doi:10.1016/S0370-2693(98)00377-3 [arXiv:hep-th/9802109 [hep-th]].

\bibitem{Witten:1998qj}
E.~Witten, ``Anti-de Sitter space and holography,'' Adv. Theor. Math. Phys. \textbf{2} (1998), 253-291 doi:10.4310/ATMP.1998.v2.n2.a2 [arXiv:hep-th/9802150 [hep-th]].

\bibitem{Jevicki:1979mb}
A.~Jevicki and B.~Sakita, ``The Quantum Collective Field Method and Its Application to the Planar Limit,'' Nucl. Phys. B \textbf{165} (1980), 511 doi:10.1016/0550-3213(80)90046-2.

\bibitem{Jevicki:1980zg}
A.~Jevicki and B.~Sakita, ``Collective Field Approach to the Large $N$ Limit: Euclidean Field Theories,'' Nucl. Phys. B \textbf{185} (1981), 89-100 doi:10.1016/0550-3213(81)90365-5

\bibitem{Vasiliev:1990en}
M.~A.~Vasiliev, ``Consistent equation for interacting gauge fields of all spins in (3+1)-dimensions,'' Phys. Lett. B \textbf{243} (1990), 378-382 doi:10.1016/0370-2693(90)91400-6.

\bibitem{Vasiliev:2003ev}
M.~A.~Vasiliev, ``Nonlinear equations for symmetric massless higher spin fields in (A)dS(d),'' Phys. Lett. B \textbf{567} (2003), 139-151 doi:10.1016/S0370-2693(03)00872-4
[arXiv:hep-th/0304049 [hep-th]].

\bibitem{Jevicki:1983hb}
A.~Jevicki and J.~P.~Rodrigues, ``Master Variables and Spectrum Equations in Large $N$ Theories,'' Nucl. Phys. B \textbf{230} (1984), 317-335 doi:10.1016/0550-3213(84)90216-5.

\bibitem{Jevicki:1993rr}
A.~Jevicki and J.~P.~Rodrigues, ``Loop space Hamiltonians and field theory of noncritical strings,'' Nucl. Phys. B \textbf{421} (1994), 278-292 doi:10.1016/0550-3213(94)90329-8
[arXiv:hep-th/9312118 [hep-th]].

\bibitem{deMelloKoch:1996mj}
R.~de Mello Koch and J.~P.~Rodrigues, ``Systematic 1/N corrections for bosonic and fermionic vector models without auxiliary fields,'' Phys. Rev. D \textbf{54} (1996), 7794-7814
doi:10.1103/PhysRevD.54.7794 [arXiv:hep-th/9605079 [hep-th]].

\bibitem{Klebanov:2002ja}
I.~R.~Klebanov and A.~M.~Polyakov, ``AdS dual of the critical O(N) vector model,'' Phys. Lett. B \textbf{550} (2002), 213-219 doi:10.1016/S0370-2693(02)02980-5 [arXiv:hep-th/0210114 [hep-th]].

\bibitem{Sezgin:2002rt}
E.~Sezgin and P.~Sundell, ``Massless higher spins and holography,'' Nucl. Phys. B \textbf{644} (2002), 303-370 [erratum: Nucl. Phys. B \textbf{660} (2003), 403-403] doi:10.1016/S0550-3213(02)00739-3 [arXiv:hep-th/0205131 [hep-th]].

\bibitem{Das:2003vw}
S.~R.~Das and A.~Jevicki, ``Large N collective fields and holography,''
Phys. Rev. D \textbf{68} (2003), 044011 doi:10.1103/PhysRevD.68.044011
[arXiv:hep-th/0304093 [hep-th]].

\bibitem{deMelloKoch:2010wdf}
R.~de Mello Koch, A.~Jevicki, K.~Jin and J.~P.~Rodrigues, ``$AdS_4/CFT_3$ Construction from Collective Fields,'' Phys. Rev. D \textbf{83} (2011), 025006 doi:10.1103/PhysRevD.83.025006 [arXiv:1008.0633 [hep-th]].

\bibitem{Jevicki:2011ss}
A.~Jevicki, K.~Jin and Q.~Ye, ``Collective Dipole Model of AdS/CFT and Higher Spin Gravity,'' J. Phys. A \textbf{44} (2011), 465402 doi:10.1088/1751-8113/44/46/465402 [arXiv:1106.3983 [hep-th]].

\bibitem{Jevicki:2011aa}
A.~Jevicki, K.~Jin and Q.~Ye, ``Bi-local Model of AdS/CFT and Higher Spin Gravity,''
[arXiv:1112.2656 [hep-th]].

\bibitem{deMelloKoch:2012vc}
R.~de Mello Koch, A.~Jevicki, K.~Jin, J.~P.~Rodrigues and Q.~Ye, ``S=1 in O(N)/HS duality,'' Class. Quant. Grav. \textbf{30} (2013), 104005 doi:10.1088/0264-9381/30/10/104005 [arXiv:1205.4117 [hep-th]].

\bibitem{deMelloKoch:2014mos}
R.~de Mello Koch, A.~Jevicki, J.~P.~Rodrigues and J.~Yoon, ``Holography as a Gauge Phenomenon in Higher Spin Duality,''
JHEP \textbf{01} (2015), 055 doi:10.1007/JHEP01(2015)055
[arXiv:1408.1255 [hep-th]].

\bibitem{deMelloKoch:2014vnt}
R.~de Mello Koch, A.~Jevicki, J.~P.~Rodrigues and J.~Yoon, ``Canonical Formulation of $O(N)$ Vector/Higher Spin Correspondence,''
J. Phys. A \textbf{48} (2015) no.10, 105403 doi:10.1088/1751-8113/48/10/105403
[arXiv:1408.4800 [hep-th]].

\bibitem{deMelloKoch:2018ivk}
R.~de Mello Koch, A.~Jevicki, K.~Suzuki and J.~Yoon, ``AdS Maps and Diagrams of Bi-local Holography,'' JHEP \textbf{03} (2019), 133 doi:10.1007/JHEP03(2019)133
[arXiv:1810.02332 [hep-th]].

\bibitem{deMelloKoch:2021cni}
R.~de Mello Koch, E.~Gandote, N.~H.~Tahiridimbisoa and H.~J.~R.~Van Zyl, ``Quantum error correction and holographic information from bilocal holography,'' JHEP \textbf{11} (2021), 192 doi:10.1007/JHEP11(2021)192 [arXiv:2106.00349 [hep-th]].

\bibitem{deMelloKoch:2022sul}
R.~de Mello Koch and G.~Kemp, ``Holography of information in AdS/CFT,'' JHEP \textbf{12} (2022), 095 doi:10.1007/JHEP12(2022)095 [arXiv:2210.11066 [hep-th]].

\bibitem{deMelloKoch:2023ngh}
R.~de Mello Koch, ``Microscopic entanglement wedges,'' JHEP \textbf{08} (2023), 056
doi:10.1007/JHEP08(2023)056 [arXiv:2307.05032 [hep-th]].

\bibitem{Aharony:2020omh}
O.~Aharony, S.~M.~Chester and E.~Y.~Urbach, ``A Derivation of AdS/CFT for Vector Models,''
JHEP \textbf{03} (2021), 208 doi:10.1007/JHEP03(2021)208
[arXiv:2011.06328 [hep-th]].

\bibitem{Aharony:2021ovo}
O.~Aharony, S.~M.~Chester and E.~Y.~Urbach, ``AdS from CFT for scalar QED,''
Phys. Rev. D \textbf{104} (2021) no.12, 126011 doi:10.1103/PhysRevD.104.126011
[arXiv:2109.05512 [hep-th]].

\bibitem{Aharony:2022feg}
O.~Aharony, S.~M.~Chester, T.~Sheaffer and E.~Y.~Urbach, ``Explicit holography for vector models at finite N, volume and temperature,''
JHEP \textbf{03} (2023), 016 doi:10.1007/JHEP03(2023)016
[arXiv:2208.13607 [hep-th]].

\bibitem{Mulokwe:2018czu}
M.~Mulokwe and J.~P.~Rodrigues, ``Large N bilocals at the infrared fixed point of the three dimensional O(N) invariant vector theory with a quartic interaction,'' JHEP \textbf{11} (2018), 047, doi:10.1007/JHEP11(2018)047 [arXiv:1808.00042 [hep-th]].

\bibitem{Johnson:2022cbe}
C.~Johnson, M.~Mulokwe and J.~P.~Rodrigues, ``Constructing the bulk at the critical point of three-dimensional large N vector theories,'' Phys. Lett. B \textbf{829} (2022), 137056
doi:10.1016/j.physletb.2022.137056 [arXiv:2201.10214 [hep-th]].

\bibitem{Fronsdal:1978rb}
C.~Fronsdal, ``Massless Fields with Integer Spin,''
Phys. Rev. D \textbf{18} (1978), 3624 doi:10.1103/PhysRevD.18.3624.

\bibitem{Das:1990kaa}
S.~R.~Das and A.~Jevicki, ``String Field Theory and Physical Interpretation of $D=1$ Strings,'' Mod. Phys. Lett. A \textbf{5} (1990), 1639-1650 doi:10.1142/S0217732390001888.

\bibitem{Demeterfi:1991tz}
K.~Demeterfi, A.~Jevicki and J.~P.~Rodrigues, ``Scattering amplitudes and loop corrections in collective string field theory,'' Nucl. Phys. B \textbf{362} (1991), 173-198 doi:10.1016/0550-3213(91)90561-B

\bibitem{Demeterfi:1991nw}
K.~Demeterfi, A.~Jevicki and J.~P.~Rodrigues, ``Scattering amplitudes and loop corrections in collective string field theory. 2.,'' Nucl. Phys. B \textbf{365} (1991), 499-519 doi:10.1016/S0550-3213(05)80030-6

\bibitem{Jevicki:1991yi}
A.~Jevicki, ``Nonperturbative collective field theory,'' Nucl. Phys. B \textbf{376} (1992), 75-98 doi:10.1016/0550-3213(92)90068-M

\bibitem{Jevicki:1982jj}
A.~Jevicki, O.~Karim, J.~P.~Rodrigues and H.~Levine, ``Loop Space Hamiltonians and Numerical Methods for Large $N$ Gauge Theories,'' Nucl. Phys. B \textbf{213} (1983), 169-188 doi:10.1016/0550-3213(83)90180-3

\bibitem{Jevicki:1983wu}
A.~Jevicki, O.~Karim, J.~P.~Rodrigues and H.~Levine, ``Loop Space Hamiltonians and Numerical Methods for Large $N$ Gauge Theories. 2.,'' Nucl. Phys. B \textbf{230} (1984), 299-316 doi:10.1016/0550-3213(84)90215-3

\bibitem{Koch:2021yeb}
R.~d.~Koch, A.~Jevicki, X.~Liu, K.~Mathaba and J.~P.~Rodrigues, ``Large N optimization for multi-matrix systems,'' JHEP \textbf{01} (2022), 168 doi:10.1007/JHEP01(2022)168
[arXiv:2108.08803 [hep-th]].

\bibitem{Mathaba:2023non}
K.~Mathaba, M.~Mulokwe and J.~P.~Rodrigues, ``Large N Master Field Optimization: the Quantum Mechanics of two Yang-Mills coupled Matrices,'' [arXiv:2306.00935 [hep-th]].

\bibitem{Metsaev:1999ui}
R.~R.~Metsaev, ``Light cone form of field dynamics in Anti-de Sitter space-time and AdS / CFT correspondence,'' Nucl. Phys. B \textbf{563} (1999), 295-348 
doi:10.1016/S0550-3213(99)00554-4 [arXiv:hep-th/9906217 [hep-th]].

\bibitem{Metsaev:2008fs}
R.~R.~Metsaev, ``Shadows, currents and AdS,''
Phys. Rev. D \textbf{78} (2008), 106010 doi:10.1103/PhysRevD.78.106010
[arXiv:0805.3472 [hep-th]].

\bibitem{Metsaev:2008ks}
R.~R.~Metsaev, ``CFT adapted gauge invariant formulation of arbitrary spin fields in AdS and modified de Donder gauge,'' Phys. Lett. B \textbf{671} (2009), 128-134
doi:10.1016/j.physletb.2008.12.002 [arXiv:0808.3945 [hep-th]].

\bibitem{Shenker:2011zf}
S.~H.~Shenker and X.~Yin, ``Vector Models in the Singlet Sector at Finite Temperature,''
[arXiv:1109.3519 [hep-th]].

\bibitem{Susskind:1998dq}
L.~Susskind and E.~Witten, ``The Holographic bound in anti-de Sitter space,'' [arXiv:hep-th/9805114 [hep-th]].


\bibitem{Corley:2001zk}
S.~Corley, A.~Jevicki and S.~Ramgoolam, ``Exact correlators of giant gravitons from dual N=4 SYM theory,'' Adv. Theor. Math. Phys. \textbf{5} (2002), 809-839 doi:10.4310/ATMP.2001.v5.n4.a6 [arXiv:hep-th/0111222 [hep-th]].

\bibitem{Brown:2007xh}
T.~W.~Brown, P.~J.~Heslop and S.~Ramgoolam, ``Diagonal multi-matrix correlators and BPS operators in N=4 SYM,'' JHEP \textbf{02} (2008), 030 doi:10.1088/1126-6708/2008/02/030 [arXiv:0711.0176 [hep-th]].

\bibitem{Bhattacharyya:2008rb}
R.~Bhattacharyya, S.~Collins and R.~de Mello Koch, ``Exact Multi-Matrix Correlators,'' JHEP \textbf{03} (2008), 044 doi:10.1088/1126-6708/2008/03/044 [arXiv:0801.2061 [hep-th]].


\end{thebibliography}
\end{document}